\documentclass[english,reprint,showpacs,preprintnumbers,nofootinbib,amsmath,amssymb,prl]{revtex4-1}
\usepackage[latin9]{inputenc}
\usepackage{array}
\usepackage{mathrsfs}
\usepackage{multirow}
\usepackage{amsmath}
\usepackage{amssymb}
\usepackage{graphicx}
\usepackage{esint}

\makeatletter

\providecommand{\tabularnewline}{\\}

\@ifundefined{textcolor}{}
{%
\definecolor{BLACK}{gray}{0}
\definecolor{WHITE}{gray}{1}
\definecolor{RED}{rgb}{1,0,0}
\definecolor{GREEN}{rgb}{0,1,0}
\definecolor{BLUE}{rgb}{0,0,1}
\definecolor{CYAN}{cmyk}{1,0,0,0}
\definecolor{MAGENTA}{cmyk}{0,1,0,0}
\definecolor{YELLOW}{cmyk}{0,0,1,0}
}

\usepackage{bm}% bold math
\usepackage{multirow}\usepackage{color}\usepackage{mathrsfs}

\usepackage{fouriernc}

\usepackage{babel}

\makeatother

\usepackage{babel}
\begin{document}
\global\long\def\figurename{FIG.}
 \addto\captionsenglish{\global\long\def\figurename{FIG.}
}

\title{Arbitrary degree distribution and high clustering in \\
 networks of locally interacting agents }

\author{Navid Dianati\textsuperscript{1} and Nima Dehmamy\textsuperscript{2}\medskip{}
 }

\affiliation{\textsuperscript{1}Lazer Lab, Northeastern University, Boston, MA
02115 }

\affiliation{\textsuperscript{1}IQSS, Harvard University, Cambridge, MA 02138}

\affiliation{\textsuperscript{2}Center for Polymer studies Boston University,
Boston, MA 02215 }
\begin{abstract}
We construct a class of network growth models based on local interactions
on a metric space, capable of producing arbitrary degree distributions
as well as a naturally high degree of clustering and assortativity
akin to certain biological networks. As a specific example, we study
the case of random-walking agents who form bonds only when they meet
at certain locations. The spatial distribution of these \textquotedblleft{}rendezvous
points\textquotedblright{} determines key characteristics of the network.
For any arbitrary degree distribution, we are able to analytically
solve for the required rendezvous point distribution.
\end{abstract}
\maketitle

\section{Introduction}

Many real-world networks are known to be scale-free and possess a
very short average network distance (the so-called small-world property)
\cite{newman2003structure,PhysRevLett.90.058701}. Traditionally,
models of scale-free networks such as the preferential attachment\textit{
}model of Barabási and Albert (BA) and small-world networks such as
Watts-Strogatz \cite{watts1998collective} focus on reproducing topological
properties of real-world networks without regard for the geometric
character of the underlying processes. However, in many real-world
networks such as social networks, nodes are embedded in a metric space
and links are generally established only if nodes make contact, either
through physical proximity or in the virtual world \cite{barthelemy_spatial_2011}.
It is therefore not surprising that in such networks, connection probabilities
should fall with increasing physical distance. Examples include phone
call \cite{krings2009urban} and scientific collaboration networks
\cite{katz1994geographical}. Indeed, a number of network growth models
have been proposed in which link formation between agents depends
on their distance in some abstract space \cite{barthelemy_spatial_2011,krioukov_hyperbolic_2010,krioukov_popularity_2012}.
The agents themselves, however, lack physical dynamics in these models.

In this paper, we propose a model capable of generating scale-free
networks based on locally interacting dynamic agents residing in a
metric space. A key feature of our model is our emphasis on the role
of certain locations in space in promoting bond formation, the same
way that the presence of meeting places such as universities, cafés,
etc. facilitates the formation of new social links. Specifically,
the agents stochastically traverse the space and form connections
only when they encounter each other at designated meeting places.
The global characteristics of the network are then determined by the
spatial distribution of these \textit{rendezvous points} (RP). When
viewed in reverse time, a different interpretation of the model is
possible. Now, rather than being meeting places, the RPs are seeds
from which independent stochastically moving agents are spawned. This
branching process is reminiscent of genetic evolution models where
new genes and the proteins they encode (represented as points in some
parameter space) are born through the \textit{duplications }and subsequent
mutations originating from an existing pool of genes \cite{pastor-satorras_evolving_2003,ohno_evolution_1970,hughes_evolution_1994,vidal_interactome_2011}.

Another feature of our model is that it produces a relatively high
global clustering coefficient akin to those observed in some biological
networks, including neuron firing correlations \cite{eguiluz2005scale}
and protein-protein interactions \cite{newman2003structure}. There
exist models capable of producing arbitrary degree distributions or
relatively high clustering \cite{{krapivsky_organization_2001},holme2002growing,{dorogovtsev_evolution_2000},{klemm_highly_2002}},
but BA-like models generally have low clustering unless substantially
modified \cite{volz2004random}.

The framework we introduce here is very general. The model may be
solved for agents moving according to a variety of different stochastic
processes. For any such process, given any desired degree distribution,
we can analytically solve the spatial rendezvous point distribution
that results in that degree distribution. In this paper, we first
demonstrate the procedure for a concrete example---namely, agents
moving according to an isotropic random walk---and then discuss the
general case.

\section{Network of interacting random walkers}

Consider a flat 2D space with area $V=L^{2}$. Place $N\gg1$ random
walkers uniformly at random in this space. For simplicity we will
work in units where $\frac{N}{V}\to1.$ Let $\phi_{i}(x,t)$ denote
the probability density of finding random walker $i$ at point $x$
and time $t$. Thus, without any interaction the Fokker-Planck equation
is the sourced diffusion, or heat equation 
\begin{equation}
(\partial_{t}-\nabla_{x}^{2})\phi_{i}(x,t)=J_{i}(x,t)\label{eq:FP}
\end{equation}
where we require agent $i$ to begin its walk at time $t=t_{0}$ and
position $x_{_{i}}$ by setting the source term to 
\begin{equation}
J_{i}(x,t)\equiv\delta(t-t_{0})\delta^{2}(x-x_{i}).
\end{equation}

With this source, the solutions of \eqref{eq:FP} are in fact the
retarded Green's functions $\phi_{i}(x,t)=G(x,t;x_{i},t_{0})$. In
the case of 2D diffusion, this is given by 
\begin{equation}
G(x,t_{x};y,t_{y})=\frac{\theta(t_{x}-t_{y})}{4\pi(t_{x}-t_{y})}\exp\left[-\frac{|x-y|^{2}}{4(t_{x}-t_{y})}\right].\label{eq:green}
\end{equation}
Defining the operator $\mathscr{L}_{x,t}\equiv\partial_{t}-\nabla_{x}^{2}$
and its conjugate $\mathscr{L}_{x,t}^{\dagger}=-\partial_{t}-\nabla_{x}^{2}$
we have 
\begin{align}
\mathscr{L}_{x,t}G_{i}(x,t;y,s) & =\mathscr{L}_{y,s}^{\dagger}G_{i}(x,t;y,s)\\
 & =\delta(t-s)\delta^{2}(x-y)
\end{align}

We assume that bonds are formed between two agents only when they
meet at designated locations (coffee-shops, universities, work place,
etc) in space, which we call ``rendezvous points'' or RP's, characterized
by a time-dependent spatial distribution $\Gamma(x,t).$ Once two
agents meet at an RP, there is a small chance $\lambda$ that they
form a bond. Therefore, to the lowest order in $\lambda,$ the probability
that agents $i$ and $j$ have become connected by time $T>t_{0}$
is given by 
\begin{align}
A_{ij}(t_{0},T)= & \lambda\int\int_{t_{0}}^{T}\mathrm{d}t\,\mathrm{d}^{2}xG_{i}(x,t;x_{i},t_{0})\\
 & \times\Gamma(x,t)G_{j}(x,t;x_{j},t_{0})+O(\lambda^{2}).\label{eq:A1}
\end{align}
The $A_{ij}$ may be interpreted either as elements of the weighted
dense adjacency matrix of the network of connections, or as bond probabilities,
in which case the matrix $A$ defines an ensemble of unweighted random
graphs.

\begin{figure*}
\begin{centering}
\parbox[b][1\totalheight][t]{0.75\textwidth}{%
\begin{tabular}{l>{\raggedright}m{3cm}>{\raggedright}m{4cm}>{\raggedright}m{3cm}}
\hline 
\noalign{\vskip0.06cm}
\multirow{2}{*}{} & \multirow{2}{3cm}{$\boldsymbol{k(r,t,T)}$ } & \multicolumn{2}{c}{$\boldsymbol{\Gamma(r,t)}^{\dagger}$ \ \ \ \ \ }\tabularnewline[0.07cm]
\cline{3-4} 
\noalign{\vskip0.06cm}
 &  & {\small{$k_{\mathrm{max}}(T-t)=4\pi(T-t)$}}  & {\small{$k_{\mathrm{max}}(T-t)=\pi c$}} \tabularnewline[0.15cm]
\hline 
\noalign{\vskip0.06cm}
 $\boldsymbol{\gamma}=1^{*}$  & $\frac{\theta(T-t)}{4\pi(T-t)}e^{\frac{r^{2}}{4(T-t)}}$  & $\delta^{2}(\vec{r})\delta(T-t)$  & \tabularnewline[0.15cm]
\hline 
\noalign{\vskip0.06cm}
$\boldsymbol{\gamma}=1$  & $k_{\mathrm{max}}e^{-\frac{\pi r^{2}}{k_{\mathrm{max}}}}$  & $4\pi e^{-\frac{\rho^{2}}{4}}$  & $\dfrac{4\left(c-r^{2}\right)}{\pi c^{3}}e^{-\frac{r^{2}}{c}}$ \tabularnewline[0.15cm]
\hline 
\noalign{\vskip0.06cm}
$\boldsymbol{\gamma}=2$  & $\dfrac{k_{\mathrm{max}}^{2}}{\pi r^{2}+k_{\mathrm{max}}}$  & $32\pi\dfrac{\left(\rho^{4}+4\rho^{2}+16\right)}{\left(\rho^{2}+4\right)^{3}}$  & $4\pi\dfrac{c^{2}\left(c-r^{2}\right)}{\left(2r^{2}+c\right)^{3}}$ \tabularnewline[0.15cm]
\hline 
\noalign{\vskip0.06cm}
 $\boldsymbol{\gamma}=3$  & $\left[\dfrac{k_{\mathrm{max}}^{3}}{2\pi r^{2}+k_{\mathrm{max}}}\right]^{1/2}$  & $2^{3/2}\pi\dfrac{\left(3\rho^{4}+8\rho^{2}+16\right)}{\left(\rho^{2}+2\right)^{5/2}}$  & $4\pi\dfrac{c^{3/2}\left(c-r^{2}\right)}{\left(2r^{2}+c\right)^{5/2}}$ \tabularnewline[0.15cm]
\hline 
 {\scriptsize{$^{\dagger}(\rho\equiv r/\sqrt{T-t})$}}  &  &  & \tabularnewline
\end{tabular}\\
 \vspace{0.35cm}
}\includegraphics[height=5.55cm]{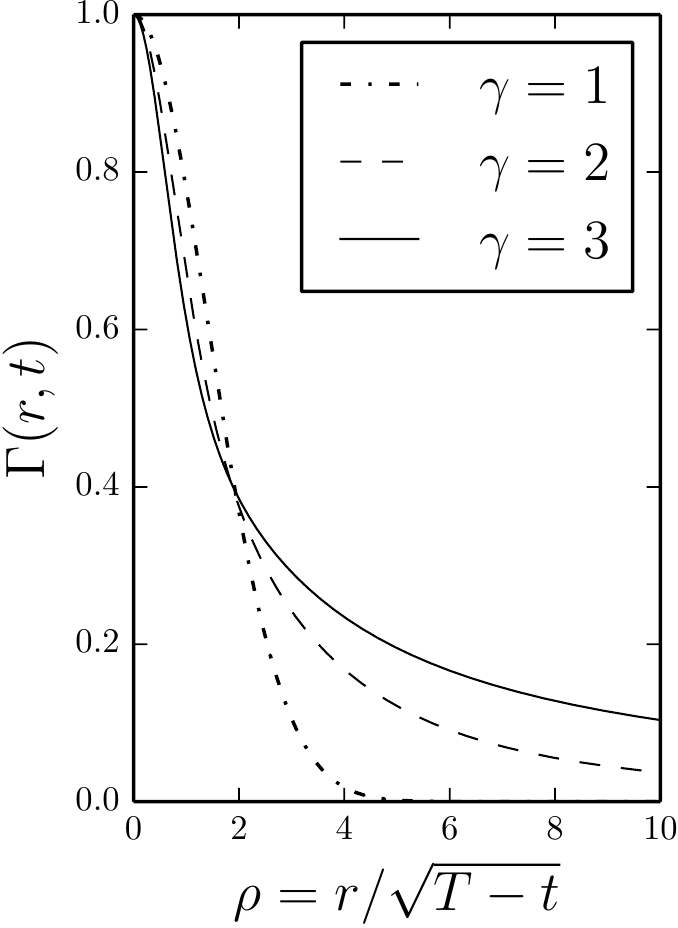} 
\par\end{centering}

\caption{The spatial degree function $k(r,t,T)$ and RP distribution function
$\Gamma(r,t)$ for various exponents in 2 spatial dimensions. The
models are characterized by $k_{\mathrm{max}}=k_{\mathrm{max}}(T-t)$
taken here to be linear: $k_{\mathrm{max}}(s)\propto s$. \label{fig:123}}
\end{figure*}

\section{Analytical results}

With this simple linear equation many network characteristics can
be computed analytically. In what follows we will first prove an important
relation between degrees and the RP distribution $\Gamma(x,t)$. Then
we will outline the procedure which allows one to 1) derive the degree
distribution when $\Gamma(x,t)$ is given, and more importantly 2)
find $\Gamma(x,t)$ such that a desired degree distribution such as
a power-law is obtained.

The degrees $k_{i}$ are defined as \ensuremath{k_{i}\equiv\sum_{j=1}^{N}A_{ij}}.
Replacing the sum with an integral in the continuum limit and using
$\int\mathrm{d}^{2}x_{j}G_{j}(x,t;x_{j},t_{0})=1$ for diffusion we
obtain 
\begin{align}
k(x_{i},t_{0},T)= & \lambda\int\int_{-\infty}^{T}\mathrm{d}t\,\mathrm{d}^{2}xG_{i}(x,t;x_{i},t_{0})\Gamma(x,t)\label{eq:k}
\end{align}
where $k(x_{i},t_{0},T)$ is the degree, measured at time $T,$ of
an agent starting at position $x_{i}$ at time $t_{0}$. Applying
$\mathscr{L}_{x_{i},t_{0}}^{\dagger}$ on both sides of \eqref{eq:k}
thus yields the first important result 
\begin{align}
\mathscr{L}_{x,t}^{\dagger}k(x,t,T)= & \lambda\theta(T-t)\Gamma(x,t)\label{eq:lkg}
\end{align}
where $\theta(x)$ is the Heaviside step function. The significance
of Eq. \eqref{eq:lkg} is in that it relates the node degrees to the
RP distribution. This allows us for instance to solve for the RP distribution
required for an arbitrary degree distribution as we now proceed to
do.

If $\Gamma$ is uniform over the space, each node will be overwhelmingly
connected to those in its close vicinity, and the translational symmetry
results in a sharply peaked degree distribution. Non-trivial degree
distributions therefore arise only when this symmetry is broken. Let
us now focus on rotationally symmetric RP distributions $\Gamma(r,t)$.
With this symmetry the degree distribution $P(k)$ is an implicit
function of $r,$ since $k$ is only a function of $r.$ For $P(k)$
monotonic (possibly with cutoffs near $k=0$ and $k_{\mathrm{max}}$),
we have 
\begin{align}
P[k(r)]|\mathrm{d}k(r)| & =|\mathrm{d}N(r)|\label{eq:pkn}\\
P[k(r)] & =\left|\frac{\mathrm{d}N}{\mathrm{d}k}\right|=\frac{\mathrm{d}N}{\mathrm{d}r}\left|\frac{\mathrm{d}k}{\mathrm{d}r}\right|^{-1}\label{eq:pknr}
\end{align}
where $\mathrm{d}N(r)$ is the number of nodes in the annulus $[r,r+\mathrm{d}r].$
The absolute value is necessary since $\mathrm{d}k/\mathrm{d}r$ may
be negative. This simple equation combined with \eqref{eq:lkg} allows
us to explicitly calculate the degree distribution given $\Gamma(x,t)$
or conversely, to solve for $\Gamma(x,t)$ given a desired degree
distribution. As a simple example, with $t_{0}=0$ and a single rendezvous
point activated at a single time, $\Gamma(r,t)=\delta(r)\delta(t-t_{e})$,
equations (\ref{eq:k}) and (\ref{eq:pknr}) yield 
\begin{align}
P[k(r)]= & 4\pi t_{e}\theta(T-t_{e})k^{-1}\label{eq:pkdelta}
\end{align}
which is a power law distribution $P(k)\propto k^{-\gamma}$ with
exponent $\gamma=1$.

\section{General power-law example}

We will now derive the conditions for $\Gamma(r,t)$ for which the
degree distribution becomes a power-law, possibly changing over time
with an overall factor $p(T-t_{0})$ and with an upper cutoff%
\footnote{The lower cutoff will depend on $L$ and $T.$%
} 
\begin{equation}
P(k;t_{0},T)=p(T-t_{0})k^{-\gamma},\quad k\in\left[1,k_{\mathrm{max}}\right].\label{eq:pow}
\end{equation}
The maximum degree $k_{\mathrm{max}}$ is chosen such that the expected
number of nodes of degree $k_{\mathrm{max}}$ is one, i.e. $P(k_{\mathrm{max}})=1$.
Therefore from \eqref{eq:pow} 
\begin{equation}
k_{\mathrm{max}}\equiv p(T-t_{0})^{1/\gamma}.\label{eq:kmax}
\end{equation}
Now, in order to solve for $\Gamma(r,t),$ we integrate \eqref{eq:pknr}
to find $k(r,t,T)$ and plug it in \eqref{eq:lkg}. We obtain 
\begin{align}
\Gamma(r,t) & =\mathscr{L}_{\vec{r},t}^{\dagger}\left[\frac{\pi(\gamma-1)r^{2}+p(T-t)^{1/\gamma}}{p(T-t)}\right]^{\frac{1}{1-\gamma}}\label{eq:powgam-1}
\end{align}
For arbitrary $\gamma>1,$ and 
\begin{align}
 & \Gamma(r,t)=\mathscr{L}_{\vec{r},t}^{\dagger}\left\{ p(T-t)\exp\left[\frac{\pi r^{2}}{p(T-t)}\right]\right\} 
\end{align}
for $\gamma=1$.

The results for $\gamma=1,2,3$ and two different $p(t)$ are given
in Fig. \ref{fig:123}. Using these results, we can simulate the model
by placing agents and RP's on a finite area of the 2D space with appropriate
distributions, and computing the $A_{ij}.$ To avoid boundary effects,
the characteristic range of the random walkers $\sigma=\sqrt{4T}$
must be much smaller than the system size $L.$ For the continuum
approximation to hold, $\sigma$ must be much larger than the inter-agent
distance $L/\sqrt{N}.$ With proper normalization, $A_{ij}$ may be
interpreted as the probability that the unweighted edge $(i,j)$ exists,
and different realizations of the network can be constructed accordingly.
\begin{figure*}[t]
\centering{}\includegraphics[width=1\textwidth]{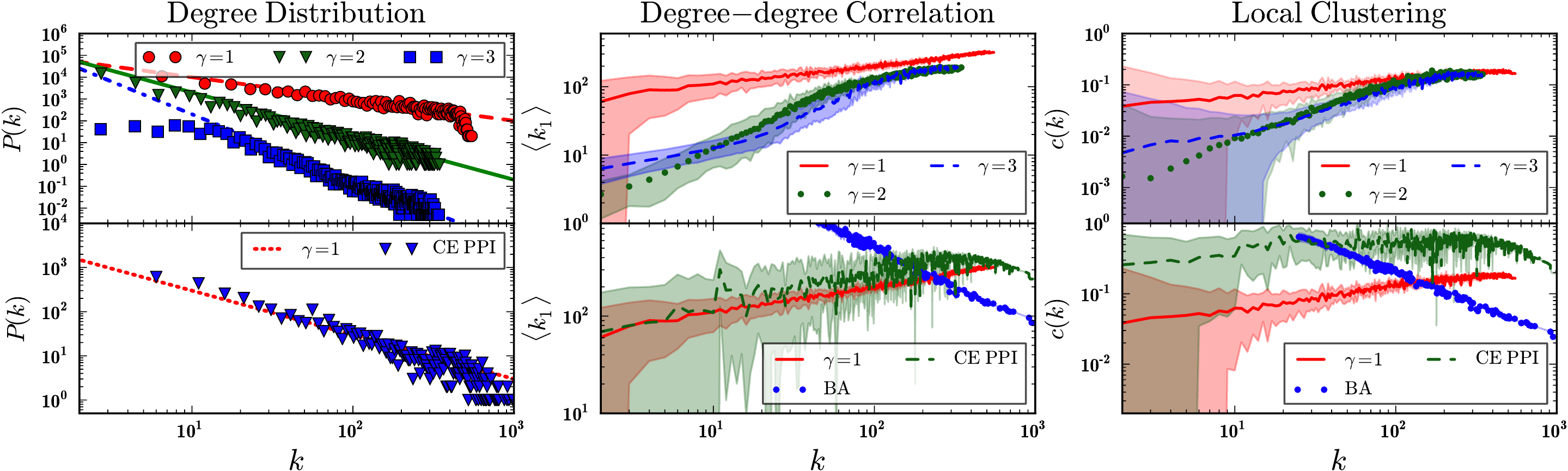}\caption{Top left: degree distributions of graphs generated from our model
for power-law distributions with $\gamma=1,2,3$. Lines represent
$k^{-\gamma}$ for $\gamma=1,2,3$ (To separate data points for clarity
a constant is multiplied into the y-axis of the 3 datasets). Top middle
and top right show average degree of first neighbors, $\langle k_{1}\rangle$
vs degrees, and local clustering $c(k)$ for the 3 simulations, respectively.
The curves are the means and the shaded area is one standard deviation
above and below mean. Lower left: degree distribution of the C. elegans
integrated protein interaction network \cite{danafarber}. $P(k)\propto k^{-1}$
seems to match it well. Lower middle compares the degree assortativity
of the C. elegans data (CE PPI) with our simulated $\gamma=1$ data,
which almost perfectly reproduces the CE PPI statistics. The blue
curve, BA, is a Barabasi-Albert network with $m=25$ (to produce comparable
density) which exhibits a different behavior from our model and most
parts of the CE PPI. Lower right compares the clustering of the three.
Our model does seem to behave similar to the CE PPI, but it is no
match for its extremely high clustering. Our global clustering for
this $\gamma=1$ is $C=0.22$ whereas the CE PPI has $C=0.47$. Number
of nodes and edges are $6.2\times10^{3}$ and $1.8\times10^{5}$ in
CE PPI and $5.4\times10^{3}$ and $2.1\times10^{5}$ in $\gamma=1$.
\label{fig:Big2x3}}
\end{figure*}

\subsection{Higher moments: assortativity and clustering}

The degree distribution is only one of many measures characterizing
a graph. It is the distribution of the first moment $k_{i}=\sum_{j}A_{ij}$.
The simplest among higher order measures of graph connectivity would
be the degree-degree correlation, also known as \textit{degree assortativity,}
which compares the degree of the first neighbor of node $i$, $k_{i}^{1}$,
to $k_{i}$ itself. The average first neighbor degree is 
\[
\langle k_{i}^{1}\rangle=\frac{1}{N}\sum_{j}A_{ij}k_{j}=\frac{1}{N}\sum_{j}[A^{2}]_{ij}
\]
and is thus related to the square of the adjacency matrix.

The next higher order measure which is related to $A^{3}$ is the
\textit{global clustering coefficient $C$ }which measures the degree
to which the graph is clustered \cite{newman2003structure} 
\begin{equation}
C\equiv\frac{3\times\#\mbox{ triangles}}{\#\mbox{ connected triplets}}.
\end{equation}
which can be shown to be equal to $C=\mathrm{Tr}\left[A^{3}\right]/\sum_{ij}\left[A^{2}\right]_{ij}$.
Clustering may also be measured at the vertex level using the \textit{local
clustering $c_{i}$} \cite{newman2003structure} defined as the number
of triangles involving node $i$ divided by the total number of such
triangles possible given the degree $k_{i}$ 
\begin{equation}
c_{i}\equiv\frac{2\times\mbox{\# triangles containing }i}{k_{i}(k_{i}-1)}.
\end{equation}
By definition $c_{i}\leq1$. Fig. \ref{fig:Big2x3} summarizes the
results of simulations for scale-free distributions with $\gamma=1,2,3.$
For each case, one realization of the unweighted random graph ensemble
is generated and the degree distributions $P(k)$, first neighbor
degree-degree correlation $\langle k^{1}\rangle$, and local clustering
$c(k)$ is shown. Interestingly, our model has a naturally high global
clustering coefficient %egree of clustering
because agents close to the RP's are all likely to connect and form
close-knit subgraphs. Fig. \ref{fig:Big2x3} illustrates how our model
compares to a particular real world network, namely the network of
protein-protein interactions in the nematode C. elegans (CE PPI) from
the integrated dataset of different types of interactions (incorporating
WI8, literature curated, Microarray, Phenotype, Interolog, and Genetic
interactions) \cite{danafarber} %compiled from the human interactome database \cite{PPI}
, as well as a Barabasi-Albert (BA) network of similar size as the
real data. The CE PPI network has a power-law degree distribution
with power $P(k)\approx k^{-1}$. We therefore compare it with a $\gamma=1$
from our model. The PPI network has an average global clustering of
$C=0.47$ versus our model's $C=0.22$. %For the BA network on the other hand, $C=0.006$.
The BA (with the same number of nodes as PPI and with $m=25$ to produce
similar density) on the other hand, has $C=0.03$ and deviates significantly
from the PPI data. %The shaded area is one standard deviation and the thick curves are the means. Our model, though having on average a smaller degree of clustering than the PPI data, follows the PPI data closely and stays within one standard deviation.
In the first two moments, $P(k)$ and $\langle k^{1}\rangle$, our
model matches the CE PPI almost perfectly. For clustering, our model
exhibits a similar trend, but falls short in terms of magnitude.

\section{Discussion}

We showed that networks with fat-tailed degree distributions and long
range connections (scale-free networks are known to be ``ultra small-world''
\cite{PhysRevLett.90.058701}) can arise from local interactions,
if the translation symmetry is broken. The framework we introduced
here uses the familiar tools of classical field theory. One of our
main results is that given any (monotonic) degree distribution, we
can analytically compute the RP distribution resulting in a network
with that degree distribution.

While we demonstrated the derivations in the case of power-law distributions,
other monotonic distributions can also be handled similarly. Furthermore,
our model is generalizable to agent dynamics other than isotropic
random walks, so long as the dynamics obeys %. In principle, one can solve the model for any agent
%dynamics with
a linear Fokker-Planck equation of the form $\mathscr{L}_{x,t}\phi(x,t)=J(x,t)$.
%where the linear operator $\mathscr{L}_{x,t}$ admits a well-defined Green's function.
Finally, the model can be solved in higher spatial dimensions as well,
with similar results.

From the point of view of application, some real world networks, especially
biological networks such as neuron firing correlation networks from
fMRI measurements \cite{eguiluz2005scale} and protein interaction
networks \cite{PPI} tend to have high global clustering coefficients
($C>20\%$). This is where many other scale-free network models such
as Barabási-Albert (BA) fall short and there have been many attempts
to remedy this \cite{holme2002growing,volz2004random,serrano2005tuning}.
One attractive feature of our model is that it has a naturally high
global clustering. It also exhibits a degree-degree correlation pattern
similar to biological data.

It must be stressed that this model was not originally intended as
a model of protein-protein interactions. Nevertheless, it contains
important elements that might constitute the ingredients for such
a model. Accumulating mutations may be conceived of as a random walk
inside some parameter space. A core set of existing genes can be represented
by a distribution $\Gamma(r,t)$. Genotypic diversification mechanisms
such as gene duplication may then correspond to branching processes,
a simple example of which we presented in our model. These elements
together with the partial empirical success of the model point to
its potential utility as a starting point for modeling biological
networks.

\section{Acknowledgement}

The authors wish to thank Shlomo Havlin for fruitful discussions.
N.D. also wishes to thank Dina S. Ghiassian, Marc Santolini and the
Barabási Lab for insightful discussions and for sharing their curated
human interactome data. This material is based upon work supported
in part by the National Science Foundation under grants No. 502019
and CMMI 1125290, and DTRA (Grant HDTRA-1-10-1-0014, Grant HDTRA-1-09-1-0035).
N.D. also thanks Mary J. Hanf for valuable insights.

N.D. and N.D. contributed equally to this work.

\bibliographystyle{plain}
\bibliography{mybib,network-growth,Biological-networks,Networks}

\end{document}